\documentclass[%
 reprint,
superscriptaddress,
showpacs,
 amsmath,amssymb,
 aps,pra,
]{revtex4-1}

\usepackage{braket}
\usepackage{graphicx}
\usepackage{dcolumn}
\usepackage{bm}
\usepackage{epstopdf}

\usepackage[colorlinks,
linkcolor=blue,
anchorcolor=blue,
urlcolor=blue,
citecolor=blue]{hyperref}

\begin{document}

\title{Entanglement Dynamics of two Non-Hermitian Qubits}

\author{Yi-Xi Zhang}
\author{Zhen-Tao Zhang}
\email{zhzhentao@163.com}%
\author{Xiao-Zhi Wei}
\author{Bao-Long Liang}
\affiliation{School of Physics Science and Information Technology, Shandong Key Laboratory of Optical Communication Science and Technology, Liaocheng University, Liaocheng 252059, China}
\author{Feng Mei}

\affiliation{State Key Laboratory of Quantum Optics and Quantum Optics Devices,
Institute of Laser Spectroscopy, Shanxi University, Taiyuan, Shanxi 030006, China}
\affiliation{Collaborative Innovation Center of Extreme Optics,
Shanxi University, Taiyuan, Shanxi 030006, China
}%
\author{Zhen-Shan Yang}
\email{yangzhenshan@lcu.edu.cn}%
\affiliation{School of Physics Science and Information Technology, Shandong Key Laboratory of Optical Communication Science and Technology, Liaocheng University, Liaocheng 252059, China}
\date{\today}

\begin{abstract}
The evolution of entanglement in a non-Hermitian quantum system may behave differently compared to its Hermitian counterpart. In this paper, we  investigate the entanglement dynamics of two coupled and driven non-Hermitian qubits. Through calculating the concurrence of the system, we find that the evolution of the bipartite entanglement manifests two distinct patterns in the parameter space. In the low non-Hermiticity regime, the concurrence oscillates significantly, while in the opposite regime the same quantity would trend to a stable value. We attribute this phenomenon to parity-time ($ \mathcal{PT}$) symmetry phase transition. In addition, we have also studied the effect of decoherence on the entanglement dynamics. Our research  provides a method to stabilize entanglement by exploiting non-Hermiticity.  

\end{abstract}

\maketitle

\section{Introduction}

Non-Hermitian quantum system is one of the hottest topics in quantum community \cite{cm1998real,z2018topological,fk2018Biorthogonal,kk2021Topological}. In contrast to Hermitian systems, quantum systems with non-Hermitian Hamiltonian have singular energy levels structures, featuring exceptional points (EPs), where two or more energy eigenstates of the Hamiltonian coalesce \cite{jz2018a,ma2019Exceptional,minganti2019quantum}. These points are usually accompanied by $ \mathcal{PT}$ symmetry phase transition \cite{sk2019Parity}. The unconventional properties of non-Hermitian systems around EPs possess important applications in quantum information processing and precision measurement \cite{bp2014loss,tg2015Observation,ld2021Experimental,pm2022Engineered}. Actually, there are several approaches to realize non-Hermitian quantum systems: introducing a mode-selective loss \cite{klauck2019observation}, embedding the desired non-Hermitian Hamiltonian into a larger Hermitian system \cite{liu2021dynamically}, removing the events of quantum jumps in a dissipative quantum system through measurement postselection\,\cite{abbasi2022topological}. In experiment, $ \mathcal{PT}$ symmetry phase transition has been observed in various quantum systems, such as ultracold atoms  \cite{dw2018Topological,Li19,ll2020Topological}, single spin  \cite{Wu19,hy2022Quantum,yw2022Tailoring} and superconducting circuit  \cite{naghiloo2019quantum,yz2023Simulation}. \\
\indent Recently, quantum entanglement in non-Hermitian systems has drawn increasing attentions \cite{py2020Entanglement,sg2021Entanglement,va2021Spreading,co2022Polarization,lm2022Quantum}. There are two research directions in this field. In the first direction, entanglement scaling with size and  phase transition were studied in many-body non-Hermition systems \cite{,AJ2014Quantum,te2014Entanglement,cg2016Equilibration,xt2023Entanglement}. In the other one, entanglement dynamics were investigated in few-body non-Hermition quantum system \cite{ak2022Maximal,yl2022Entanglement}. Here, we focus on the entanglement dynamics of two coupled non-Hermitian qubits. It was shown that the bipartite entanglement of two qubits can be increased under the local $ \mathcal{PT}$-symmetric operation on one qubit while the other qubit is Hermitian  \cite{Chen14,wang2018enhancing,grimaudo2020two}. Regarding two coupled non-Hermitian qubits, Ref. \cite{li2022speeding}  demonstrated that the maximally entanglement state can be achieved more efficiently than the Hermitian case with the same qubit coupling. However, in their scheme the entanglement is oscillating largely and the coupling strength between qubits  is perturbatively weak. It is natural to ask if it is possible to stabilize the entanglement of a quantum system by utilizing its non-Hermiticity. \\
\indent To address this question, we  theoretically investigate the entanglement dynamics of two coupled and driven non-Hermitian qubits. With fixed inter-qubit coupling and driving, we find there exist two kinds of entanglement dynamics as tuning dissipation rate of the qubits. In low dissipation regime, the concurrence used to measure the entanglement oscillates significantly with time. As the dissipation rate exceed a threshold, the concurrence would reach a stable value after several weak oscillations. To interpret this phenomenon, we have calculated the eigenvalues and eigenstates of the non-Hermitian Hamiltonian. We find that the transition of the entanglement dynamics is associated with $ \mathcal{PT}$ symmetry phase transition. In the $\mathcal{PT}$ symmetry broken phase, the concurrence of the steady state of the system is dependent on the dissipation rate: stronger dissipation yields lower concurrence of the steady state. In addition, we  also calculate and analyze the effect of decoherences of the qubits on the non-Hermitian entanglement dynamics. \\                                            
\indent This paper is organized as follows. In Sec. \ref{sc2}, we describe the construction of the model of two non-Hermitian qubits system, and present its Hamiltonian. In  Sec. \ref{sc3}, we first study the dynamic properties of the entanglement of the system by calculating the state evolution and the concurrence. Then, we calculate the eigenvalues and eigenstates of the Hamiltonian to find the EP, and analyze the relationship between the evolution of concurrence and the $\mathcal{PT}$ symmetry phase transition. In Sec. \ref{sc4}, we investigate the effect of decoherence of the two qubits on the entanglement dynamics. A conclusion is given in Sec. \ref{sc5}.
\begin{figure}\label{p1}
\includegraphics[width=8.5cm]{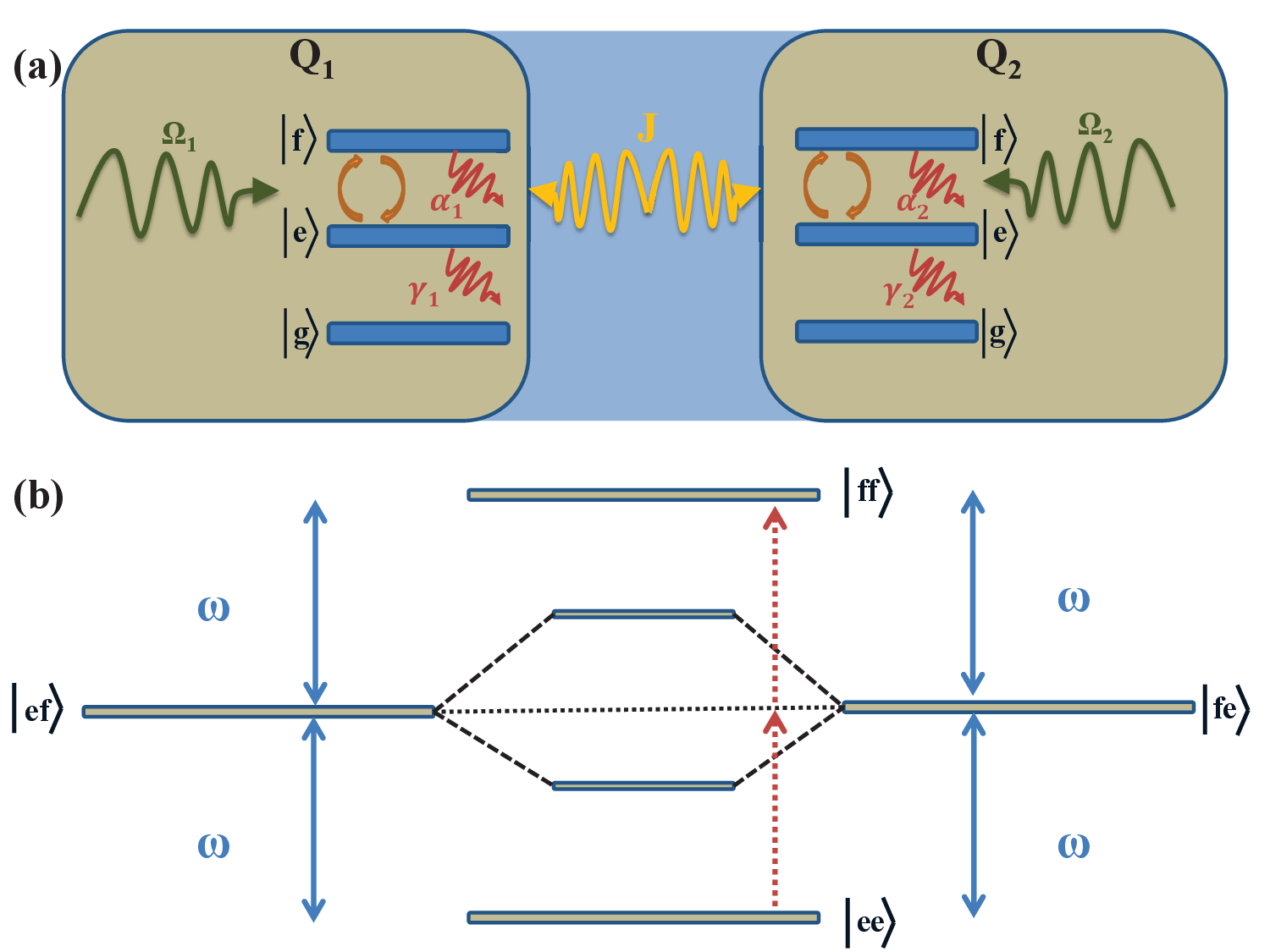}
\caption{(a) Schematic of two coupled qutrits. Each qutrit has energy levels $|g\rangle$, $|e\rangle$ and $|f\rangle$. The transition between the state $|e\rangle$ and $|f\rangle$ is driven resonantly with strength $\Omega_{1,2}$, and the state $|e\rangle$ could decay to the ground state $|g\rangle$ with rate $\gamma_{1,2}$. Besides, we denote the relaxation rate from $|f\rangle$ to $|e\rangle$ as $\alpha_{1,2}$, which are assumed to be zero before Sec. \ref{sc4}. (b) Illustration of the eigenstates of the two non-Hermitian qubits. In the absence of coupling $(J=0)$, the energy levels $|ef\rangle$ and $|fe\rangle$ are degenerate. As the coupling is turned on, the degeneracy would be lifted with a splitting $2J$. As $J>>\Omega_{1,2}$, the original resonant driving can not transition a single qubit from $g$ to $e$. However, the two-photon transition $|ee\rangle\rightarrow |ff\rangle$ is achievable because the resonant condition is satisfied. }
	\label{Fig1}
\end{figure}
\section{Model and Hamiltonian}\label{sc2}
We consider two qutrits with states of $|g\rangle$, $|e\rangle$ and $|f\rangle$ in each one, as shown in Fig. 1(a). Therein, the transition between the state $|e\rangle$ and $|f\rangle$ is driven externally, and the state $|e\rangle$ could decay to the ground state $|g\rangle$ (assume that $|f\rangle$ could not decay to $|e\rangle$, we will relax this assumption  latter). If these qutrits are prepared in the subspace $\{|e\rangle,|f\rangle\}$, we can monitor the decay of $|e\rangle$ through measuring the population of the ground state \cite{chen2021quantum}. That means the detection of $|g\rangle$ witnesses a decay from $|e\rangle$. On the contrary, the zero detection of the ground state imply the decay has not occurred actually. If we only select the zero detection cases, each qutrit is reduced to a qubit spanned on the basis $\{|e\rangle,|f\rangle\}$, whose dynamics is governed by a non-Hermitian Hamiltonian 
\begin{equation}
	H_{j=1,2}=(\Delta_j-\frac{i\gamma_j}{2})\sigma_j^+\sigma_j^-+\Omega_j\sigma_j^x , \label{eq1}
\end{equation}
where $j=1,2$ denote qubits 1 and 2, $\Delta_j=\omega_{dj}-\omega_j$ represents the detuning of the external drive relative to the qubit transition, $\gamma_j$ denotes the dissipation rate from  $|e\rangle$ state to $|g\rangle$ state, $\Omega_j$ is the driving amplitude of each qubit. The Pauli operators are defined as $\sigma_j^-=|e\rangle_j \langle f|$, $\sigma_j^+=|f\rangle_j \langle e|$, and $\sigma_j^x=\sigma_j^-+\sigma_j^+=|e\rangle_j \langle f|+|f\rangle_j\langle e|$.\\
\indent As the two qubits are coupled transversely, the total Hamiltonian reads 
\begin{equation}
H=\sum\limits_{j=1,2}H_j+J(\sigma_1^+\sigma_2^-+\sigma_1^-\sigma_2^+) , \label{Hamiltonian} 
\end{equation}
where $J$ represents the coupling strength of the two qubits. The state evolution of the system follows Schr\"{o}dinger equation with the Hamiltonian in Eq.(\ref{Hamiltonian}). Because of the non-Hermiticity of the Hamiltonian, the evolution is no longer unitary, i.e., the yielded state $|\psi\rangle$ does not preserve the unity probability. We can manually normalize the state to get  $|\tilde \psi\rangle= \frac{|\psi\rangle}{\mid |\psi\rangle \mid}=\alpha|ff\rangle+\beta|fe\rangle+\zeta|ef\rangle+\eta|ee\rangle$.\\
\indent To measure the bipartite entanglement, we introduce the quantity concurrence. For the pure state $|\tilde \psi\rangle$, the concurrence $\mathcal{C}$ can be calculated with the relation $\mathcal{C}=2\mid\alpha\eta-\beta\zeta\mid$ \cite{bennett1996concentrating}. For simplicity, we assume that both qubits have the same parameters, i.e., $\omega_1=\omega_2=\omega, \Omega_1=\Omega_2=\Omega, \gamma_1=\gamma_2=\gamma$, and consider the resonant driving: $\Delta_1=\Delta_2=0$. 
\section{ Entanglement dynamics and $ \mathcal{PT}$ symmetry phase transition }\label{sc3}
The entanglement dynamics of the non-Hermitian system are determined by a list of factors, the initial state, the coupling strength $J$, the driving $\Omega$, and the dissipation rate $\gamma$. Here, we set the initial state to be $|ff\rangle$. In this case, the driving and the coupling are both needed for generating the bipartite entanglement, and the non-Hermitian term in Eq. (\ref{eq1}) plays an important role in this process. 
\subsection{Two kinds of entanglement dynamics}
Now we investigate the evolution of the concurrence under different dissipation $\gamma$. To make the entanglement more obvious, we choose the coupling strength $J=10\, rad/\mu s$ , which is much larger than the driving amplitude $\Omega= 1.6\, rad/\mu s$. The concurrences as functions of time with various $\gamma$ are shown in Fig. 2.\\
\indent When the system is Hermitian, i.e., $\gamma=0$, the concurrence would periodically oscillate in a beating pattern. Its maximum is equal to 1, indicating a maximally entangled state turnes up. This oscillation is the consequence of the interplay of the drivings and the coupling between the qubits. In the non-Hermitian case, depending on the dissipation rate, there exist two distinct entanglement dynamics. In the low dissipation regime ($\gamma=0.5\,\mu s^{-1}$), the concurrence would also oscillate significantly, similar to the Hermitian case but with elongated period. As the dissipation rate is reaching a critical value $\gamma_c=1\mu s^{-1}$, the concurrence would not oscillate largely. Instead, it eventually stabilize around the value $\mathcal{C}=0.9$, except for weak fluctuations. This pattern is maintained on the range $\gamma> \gamma_c$, and the stabilized concurrence would descend with  increasing  $\gamma$. \\
\begin{figure} \label{p2}
	\includegraphics[width=9.0cm,height=5.0cm]{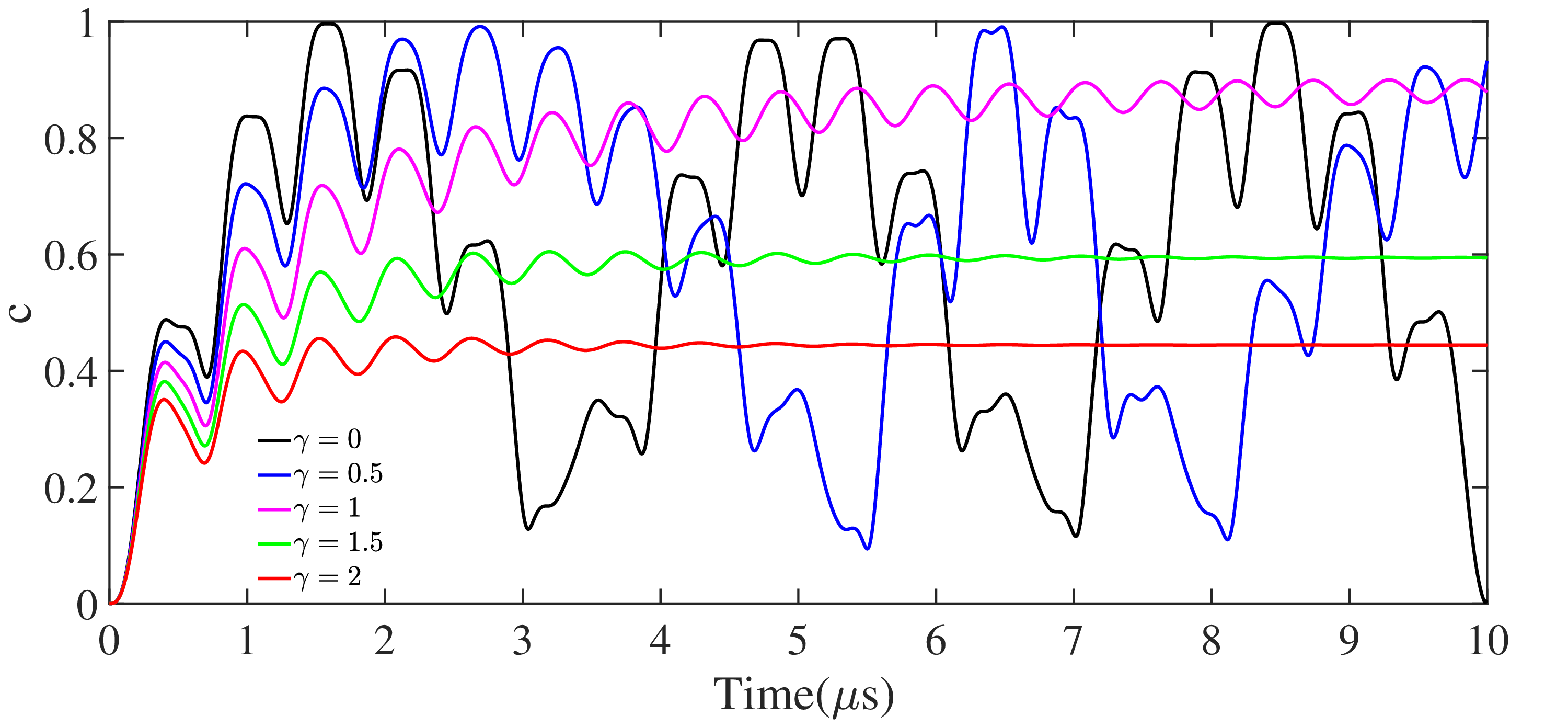}
	\caption{Concurrence evolution. The concurrence of the pure state $|\tilde \psi\rangle$ are evaluated using the relation $\mathcal{C}=2\mid\alpha\eta-\beta\zeta\mid$, with $|\tilde \psi\rangle=\alpha|ff\rangle+\beta|fe\rangle+\zeta|ef\rangle+\eta|ee\rangle$ calculated numerically from  the Schr\"{o}dinger equation. The dissipation rates are  $\gamma=0$ ( black line), $0.5\, \mu s^{-1}$ (blue line),  $1\, \mu s^{-1}$ (manganese purple line), $1.5\, \mu s^{-1} $ (green line), $2\, \mu s^{-1}$ (red line). The other parameters are set as: $J=10 \, rad/ \mu s$ , $\Omega=1.6\, rad/ \mu s$.}
\end{figure}
 To uncover the formation of the entangled state, we  calculate the state vector of the system in time for $\gamma=0.5\,\mu s^{-1}$ and $1.5\,\mu s^{-1}$. The populations of basis states $\{|ff\rangle,|ee\rangle,|fe\rangle,|ef\rangle\}$ are illustrated in Fig. 3. We can see that the system is predominantly populated in the subspace ${|ff\rangle,|ee\rangle}$ for both dissipation rates. This phenomenon can be interpreted from the structure of the energy levels shown in Fig. 1(b). Note that the initial state we consider is $|ff\rangle$. In the absence of qubit coupling, the transitions $|ee\rangle/|ff\rangle\rightarrow |ef\rangle/|ef\rangle$ can be achieved due to the resonant drivings on the two qubits. When the coupling is switched on, the degeneracy of $|ef\rangle$ and $|fe\rangle$ is lifted, and two dressed states are generated, which are superposition of the states $|ef\rangle$ and $|fe\rangle$. Due to the large energy level splitting (equal to $2J$), the single qubit driving would barely transition the system from $|ee\rangle/|ff\rangle$ to the dressed states. On the other hand, the two-photon \cite{scully1999quantum} transition $|ee\rangle\rightarrow |ff\rangle$ can be realized because the energy difference between these two levels is not affected by the coupling. Therefore, the entangled state is mainly the superposition of $|ee\rangle$ and $|ff\rangle$. Furthermore, by comparing Fig. 3 with Fig. 2, we can see that the two kinds of entanglement dynamics  reflect the population evolution of $|ee\rangle$ and $|ff\rangle$. 
\begin{figure}\label{p3}
\includegraphics[scale=0.18]{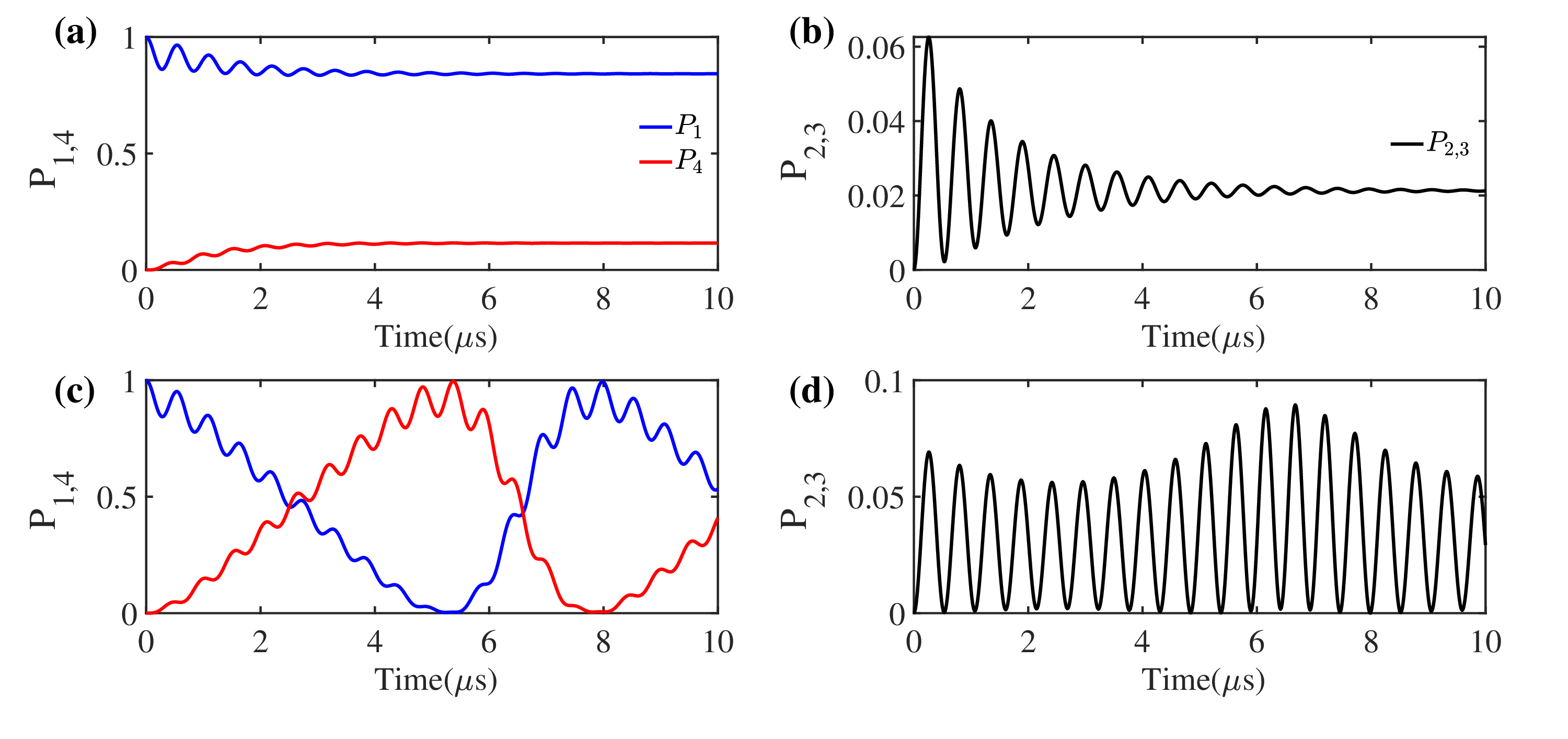}
\caption{State evolution. The populations of $|ff\rangle$, $|ee\rangle$, $|fe\rangle$, $|ef\rangle$ are plotted for $\gamma=1.5\, \mu s^{-1}$ (a-b), and $\gamma=0.5\, \mu s^{-1}$ (c-d). $P_1$, $P_2$, $P_3$ and $P_4$ represent the populations of $|ff\rangle$, $|fe\rangle$, $|ef\rangle$, $|ee\rangle$, respectively. Due to the identical parameters for the two qubits, $P_2$ is always equal to $P_3$. The other parameters are the same as Fig. 2.}
\end{figure}
\subsection{Exceptional points and phase transition }
Non-Hermitian systems feature $\mathcal{PT}$ symmetry phase transition. Now we investigate the relation between the entanglement dynamics and $ \mathcal{PT}$ symmetry in the considered system. To this end, we have calculated eigenvalues and eigenstates of the non-Hermitian Hamiltonian as a function of $\gamma$ with $J=10 \, rad/ \mu s$ and $\Omega=1.6\,rad/ \mu s$. The real and imaginary part of the eigenvalues are shown in Fig. 4 (a,b). It can be seen that at about $\gamma = 1\, \mu s^{-1}$ two eigenvalues are degenerate, where the corresponding eigenstates (not shown) coalesce as well. Thus, we know that this point is a second-order EP of the system. Moreover, as $\gamma<1\, \mu s^{-1}$, the imaginary part of the eigenvalues of the Hamiltonian are degenerate, indicating the $\mathcal{PT}$ symmetry is conserved in this region. Contrarily, as $\gamma>1\, \mu s^{-1}$, the $\mathcal{PT}$ symmetry is broken. More importantly, the EP is the exact point where the pattern of the entanglement dynamics transitions from large oscillation to stabilization. This consistently implies that the two kinds of entanglement dynamics are related with the $\mathcal{PT}$ symmetry unbroken phase and broken phase, respectively. To further verify this judgment, we  calculate and plot the EP curve and the phase diagram in the parameter space of $\Omega$ and $\gamma$, as shown in Fig. 4 (c). We find that the correlation between the entanglement dynamics and $ \mathcal{PT}$ symmetry is universal in the phase space.\\
\indent Now we elucidate the two patterns of the entanglement evolution from the view of the complex spectrum in Fig. 4. The time-dependent state vector of the system can be expressed as
\begin{equation}
	|\psi\rangle=\sum\limits_{n}{c_ne^{-itRe(\lambda_n)}e^{tIm(\lambda_n)}|\lambda_n\rangle}.
\end{equation}
where the coefficient $c_n$ is determined by the initial state, $\lambda_i$ and $|\lambda_i\rangle$ are eigenvalues and corresponding eigenstates of the Hamiltonian, respectively. It can be seen that the negative imaginary part $Im(\lambda_n)$ determines the decay of the population of the eigenstate $|\lambda_i\rangle$ in the state vector. As the system is located in the $ \mathcal{PT}$ symmetry unbroken phase, the imaginary part of all eigenstates are the same, therefore, after normalization the populations of the participated eigenstates would by no means change. Furthermore, the different real parts of the eigenvalues give rise to the oscillations of the state vector on the basis $\{|ff\rangle,|ee\rangle,|fe\rangle,|ef\rangle\}$. On the other side, if the system is in the $ \mathcal{PT}$ symmetry broken phase, the imaginary part of the eigenstates are distinct, which leads to different decay rates of the eigenstates. After a sufficiently long time, in the normalized state, the  eigenstate with lowest decay rate would survive and all the other component would fade off, resulting in a steady state at last.\\
\indent It is worth noting that the non-Hermiticity is only effective in the subspace $\{|ff\rangle,|ee\rangle\}$. In the remaining subspace, 
 $|fe\rangle$ and $|ef\rangle\}$ could decay with a common rate $\gamma$ to $|fg\rangle$, $|gf\rangle\}$ respectively. Henceforth, the non-Hermitian terms of the Hamiltonian in these two dimensions are equal and do not contribute to the dynamics in the subspace. Consequently, if the initial state is in this subspace, there is no $ \mathcal{PT}$ symmetry phase transition unless $\gamma_1\neq\gamma_2$. This can be seen in Fig. 4(b), where a couple of eigenvalues of the Hamiltonian are always degenerate on the whole range of $\gamma$. That is the reason why we choose the initial state $|ff\rangle$.
\begin{figure}\label{p4}
\includegraphics[scale=0.17]{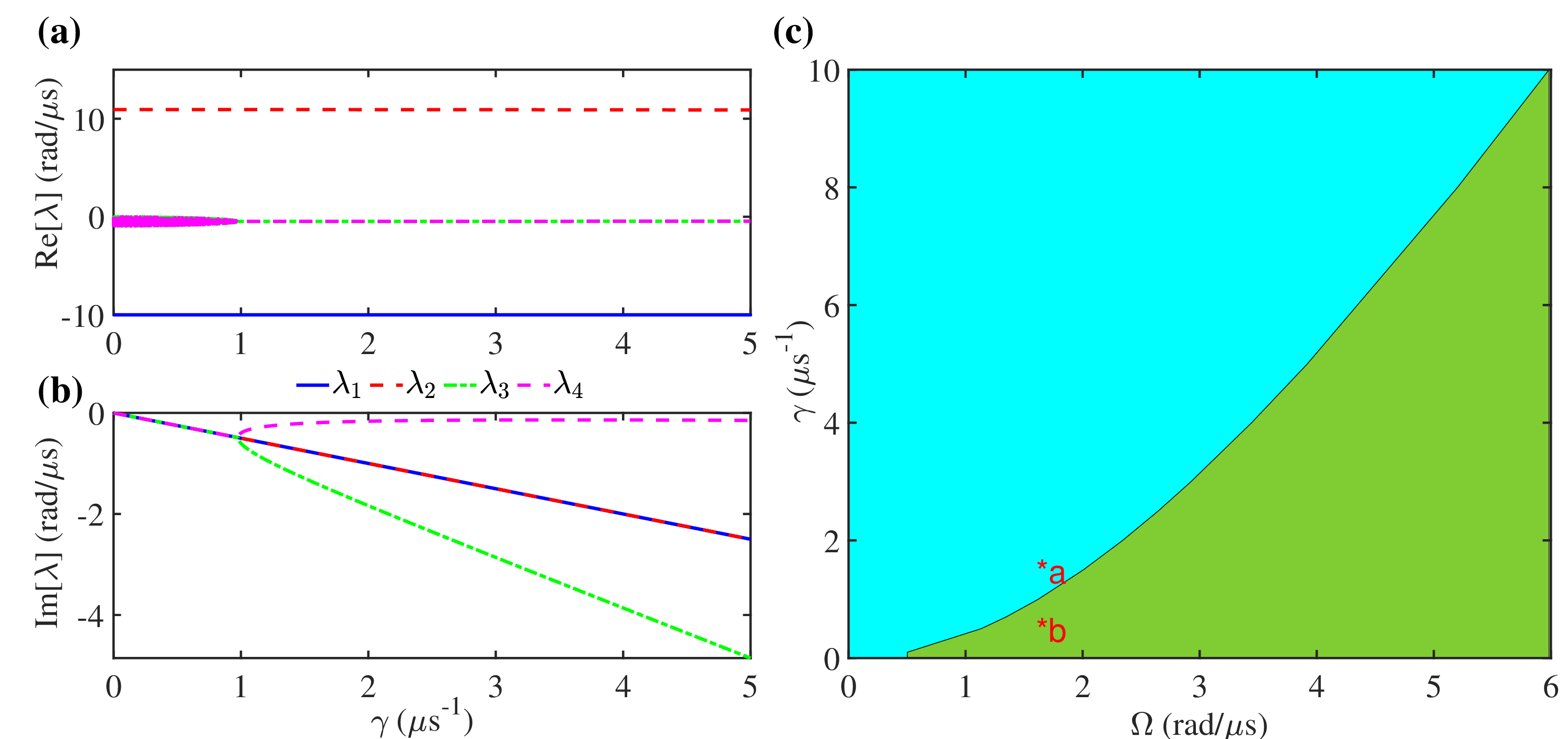}
\caption{ Complex energy spectrum and phase diagram. The real part (a) and imaginary  part (b) of the eigenvalues of the Hamiltonian are illustrated as functions of $\gamma$ with $J=10\, rad/\mu s $ and $\Omega=1.6\,rad/\mu s$. (c) phase diagram in the parameters space of $\Omega$ and $\gamma$. The upper-left (blue) region is $\mathcal{PT}$ symmetry broken phase, and the down-right region is $\mathcal{PT}$ symmetry unbroken phase. The boundary between them is the line of EP. At the point a, the state evolution has been plotted in Fig. 3(a)-(b), while the state evolution at the point b has been shown in Fig. 3(c)-(d).}
\end{figure}

\section{Influence of Decoherence on Entanglement}\label{sc4}
In the previous sections, we have not taken account of the relaxation from $|f\rangle$ to $|e\rangle$. In practice, this decoherence channel is always on. Now we consider its effect on the entanglement dynamics of the non-Hermitian system. Suppose that the relaxation rate $\alpha_{1,2}$ of $|f\rangle$ for the two qubits are equal, i.e., $\alpha_1=\alpha_2=\alpha$. Under the decoherence, the system would be in a mixed state, described by density matrix $\hat{\rho}$. Conditioned on no quantum jump from $|e\rangle$ to $|g\rangle$ and without monitoring the relaxation from $|f\rangle$ to $|e\rangle$, the evolution of $\hat{\rho}$ is governed by the Lindblad master equation  \cite{minganti2019quantum,lidar2019lecture}
\begin{equation}
	\partial_t\hat{\rho}(t)=-i[\hat{H}\hat{\rho}(t)-\hat{\rho}(t)\hat{H}^{\dagger}]+\sum_{j=1,2}\mathcal{D}[\hat{\Gamma}_{j}]\hat{\rho}(t),
\end{equation}
where $\hat{\Gamma}_j=\sqrt{\alpha}|e\rangle_j\langle f|$ is the jump operator for qubit $j$, and the dissipation term is 
\begin{equation}
	\mathcal{D}[\hat{\Gamma}_{j}]\hat{\rho}(t)=\hat{\Gamma}_{j} \hat{\rho}(t) \hat{\Gamma}_{j}^{\dagger}-\frac{1}{2}[\hat{\Gamma}_{j}^{\dagger} \hat{\Gamma}_{j} \hat{\rho}(t)+\hat{\rho}(t) \hat{\Gamma}_{j}^{\dagger} \hat{\Gamma}_{j} ].	
\end{equation}
The density matrix $\hat{\rho}(t)$ can be obtained by solving the above master equation. Due to the non-Hermiticity of the Hamiltonian $H$, $\hat{\rho}(t)$ is not trace preserving. Actually, the trace of $\hat{\rho}(t)$ denotes the probability of no quantum jump from $|e\rangle$ until $t$
\begin{equation}
	P=\text{Tr}[\hat{\rho}(t)].
\end{equation}
In experiment, $P$ represents the success probability to obtain the non-Hermitian system by post-selection \cite{chen2022decoherence}. To demonstrate the impact of decoherence on this quantity, we calculate $P$ in time for various $\alpha$ (shown in Fig. 5(a)). As expected, $P$ decrease monotonously from 1 to 0 for any fixed decoherence rate $\alpha$. With the increasing of $\alpha$, $P$ decays more quickly, which makes the observation of non-Hermitian effects more difficult. The reason lies at that the relaxation process $|f\rangle\rightarrow|e\rangle$ enhances the population of $|e\rangle$, which further increases the probability of quantum jumps from $|e\rangle$ to $|g\rangle$ .\\
\indent On the other hand, the decoherence would affect the state evolution of the system, and thus its entanglement. Before evaluating concurrences, we first normalize the density matrix through $\tilde{\hat{\rho}}=\hat{\rho}/P$. For the mixed state $\tilde{\hat{\rho}}$, the concurrence is defined as \cite{wk1998Entanglement}:
\begin{equation}
	\mathcal{C}(\tilde{\hat{\rho}})=Max\{\sqrt{\Lambda_1}-\sqrt{\Lambda_2}-\sqrt{\Lambda_3}-\sqrt{\Lambda_4},0\},
\end{equation}
where $\Lambda_{1,2,3,4}$ are the four descending eigenvalues of $\tilde{\hat{\rho}}(\sigma^y\otimes\sigma^y)\tilde{\hat{\rho}}^\ast(\sigma^y\otimes\sigma^y)$. Here, $\sigma^y$ is Pauli-y matrix,  $\tilde{\hat{\rho}}^\ast$ is the complex conjugate of $\tilde{\hat{\rho}}$.
Figure 5(b) shows the evolution of the concurrence in the presence of the decoherence. The coupling and driving strength are the same as those in the former section. As $\alpha=0.1\, \mu s^{-1}$, the two entanglement patterns are still discernible, although the concurrence are suppressed overall for both phases. As the decoherence rate goes up to $0.5\,\mu s^{-1}$, no remarkable entangled state can be produced whether $\gamma$ is smaller or larger than the critical value $\gamma_c=1\,\mu s^{-1}$. This is because the strong decoherence leads the system to a mixed state involving all the basis states. \\
\indent In a word, the relaxation of the qubits is harmful to observe the two pattern entanglement dynamics of the non-Hermitian system. However, if the relaxation rates of both qubits are low enough, it is totally possible to demonstrate the phase transition associated with bipartite entanglement in experiment. 
\begin{figure}\label{p5}
\includegraphics[scale=0.175]{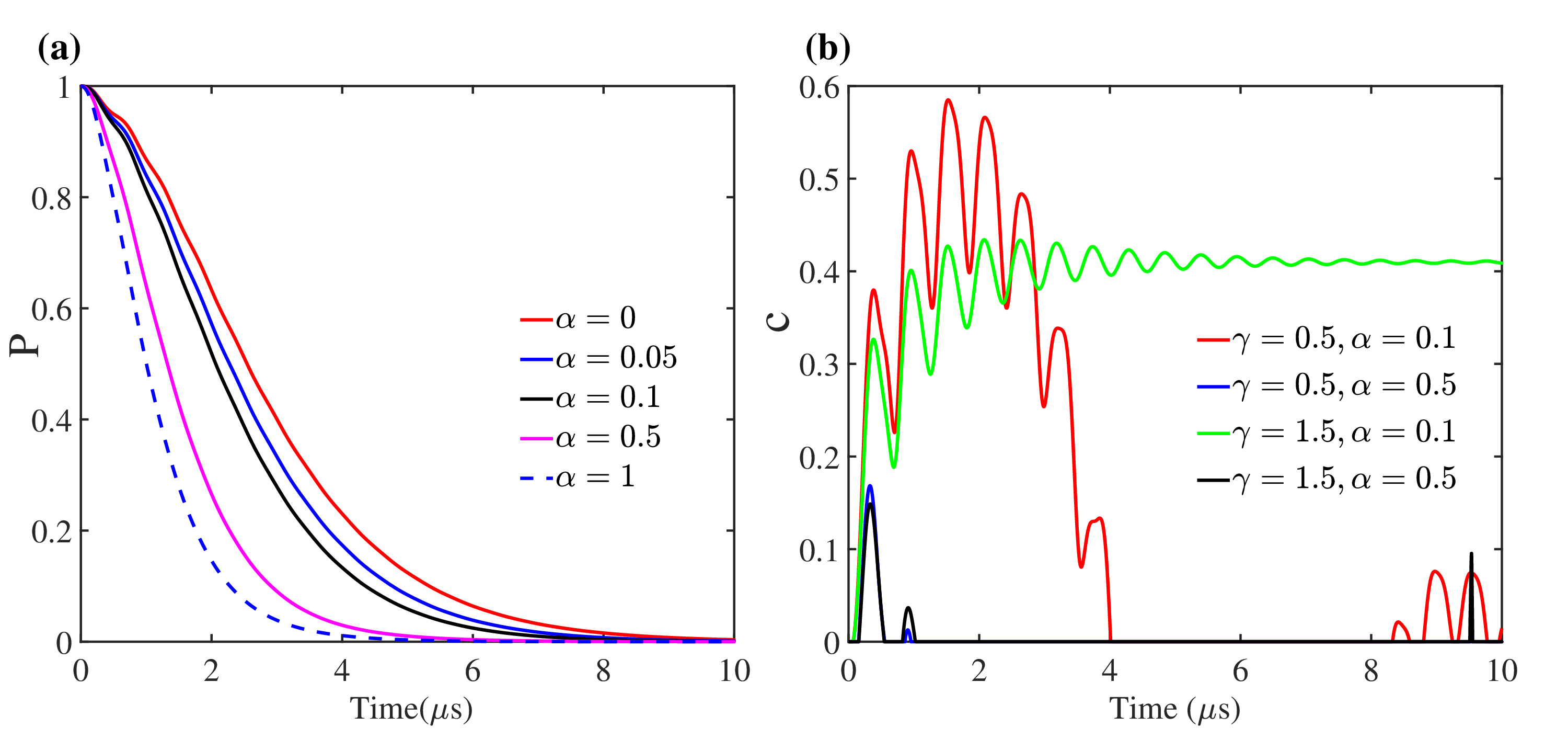}
\caption{Effects of  decoherence. (a) Probability of no quantum jump. The probability $P$ in time is shown with $\gamma=1\,\mu s^{-1}$ for $\alpha= 0,\, 0.05,\, 0.1,\, 0.5,\, 1\,\mu s^{-1}$ ( lines from right to left). (b) Concurrence in the presence of decoherence with $\gamma=0.5\,\mu s^{-1}, \alpha=0.1$ (red line), $0.5$ (blue line) $\mu s^{-1}$, and $\gamma=1.5\,\mu s^{-1}, \alpha=0.1$ (green line), $0.5$ (black line) $\mu s^{-1}$.  The other parameters are set as: $J=10 \, rad/ \mu s$ , $\Omega=1.6\, rad/ \mu s$.}
\end{figure}
\section{Conclusion }\label{sc5}
\indent In summary, we have investigated the entanglement dynamics of two coupled non-Hermitian qubits. The quantity of concurrence used to measure the entanglement has been calculated through solving the Schr\"{o}dinger equation. Starting with a separable state, the evolution of the bipartite entanglement manifests two distinct patterns in the parameter space. In the low non-Hermiticity regime, the concurrence oscillates significantly, while in the opposite regime this quantity would trend to a stable value. Through evaluating the eigenvalues and eigenstates of the system, we find that these two entanglement dynamics  correspond to $ \mathcal{PT}$ symmetry unbroken and broken phases. Also, we have  studied the effect of decoherence on the entanglement dynamics. The result shows that the decoherence may spoil the generation of entanglement in this non-Hermitian system, and thus make these two entanglement dynamics disappear. However, if the decoherence rate is small enough, these two kinds of entanglement dynamics can still be observed in experiment. Our research provides a method to stabilize entanglement by employing non-Hermiticity, which could be applied to quantum computation and quantum precision measurement.  \\
\indent The present work have assumed the two qubits possess the identical non-Hermiticity and driving. It is interesting and still unclear how the entanglement of two different non-Hermitian qubits evolves in time. We leave this question to  future studies. 

\section{Acknowledgments}
This work was funded by Natural Science Foundations of Shandong Province of China (Grant No. ZR2021MA091), Introduction and Cultivation Plan of Youth Innovation Talents for Universities of Shandong Province (Research and Innovation Team on Materials Modification and Optoelectronic Devices at extreme conditions).

\end{document}